\documentclass[11pt,letterpaper]{article}
\usepackage{systeme} 
\usepackage[utf8]{inputenc}
\usepackage{multirow} 
\usepackage{booktabs} 
\usepackage[english]{babel}
\usepackage{amsmath}
\usepackage{amsfonts}
\usepackage{amssymb}
\usepackage{graphicx}
\usepackage[small,bf]{caption} 
\usepackage[left=3cm,right=3cm,top=2cm,bottom=3cm]{geometry}
\usepackage{framed} 
\usepackage{color} 
\usepackage{wrapfig}\definecolor{shadecolor}{RGB}{220,220,220} 
\usepackage{float} 
\usepackage{array} 
\usepackage{caption} 

\author{Jorge Pinochet}
\title{\textbf{Black holes and bits: A simple path to Bekenstein-Hawking entropy}}
\begin{document}

\author{Jorge Pinochet$^{*}$\\ \\
 \small{$^{*}$\textit{Facultad de Ciencias Básicas, Departamento de Física. }}\\
  \small{\textit{Centro de Desarrollo de Investigación CEDI-UMCE,}}\\
 \small{\textit{Universidad Metropolitana de Ciencias de la Educación,}}\\
 \small{\textit{Av. José Pedro Alessandri 774, Ñuñoa, Santiago, Chile.}}\\
 \small{e-mail: jorge.pinochet@umce.cl}\\}

\date{}
\maketitle

\begin{center}\rule{0.9\textwidth}{0.1mm} \end{center}
\begin{abstract}
\noindent In the early 1970s, Jacob Bekenstein discovered that black holes have entropy, which became one of the greatest scientific revolutions of the second half of the 20th century. The objective of this paper is to present a simple derivation —partly heuristic and partly geometric— of the equation for the entropy of a black hole, which we now know as the Bekenstein-Hawking entropy. We will also briefly explore the physical implications of this equation and its relationship to the work of Stephen Hawking.\\ \\


\begin{center}\rule{0.9\textwidth}{0.1mm} \end{center}
\end{abstract}

\maketitle

\section{Introduction}
In the early 1970s, a young and introverted physics PhD student named Jacob Bekenstein discovered a connection between gravity, quantum theory, and thermodynamics that became one of the greatest scientific revolutions of the second half of the 20th century. What Bekenstein discovered was that black holes have entropy, which implies that they possess a large number of internal configurations that make them extremely complex [1], contradicting the prevailing scientific consensus of the 1970s, which established that these objects are simple, since they can be described by only three classical, externally observable parameters: mass, angular momentum and electric charge. Shortly thereafter, Bekenstein's results became the starting point for the brilliant work that led Stephen Hawking to demonstrate that black holes are not so black, since they have temperature, emit thermal radiation, and gradually evaporate [2,3].\\

The objective of this paper is to present a simple derivation, half heuristic and half geometric, of the equation for the entropy of a black hole, which we now know as the Bekenstein-Hawking (BH) entropy. We will also briefly explore the physical implications of this equation and its relationship to Hawking's pioneering work. Readers interested in further exploration of the topics covered will find accessible expositions, for example, in [4–9].

\section{Bekenstein-Hawking Entropy}

The seminal work of Bekenstein and Hawking applies to the simplest type of black hole, known as a \textit{Schwarzschild black hole}, so we will henceforth focus our analysis on these objects. A Schwarzschild black hole is completely characterized by a single physical parameter: its mass. Fig. 1 illustrates the structure of this object. Its entire mass is concentrated in a \textit{central singularity} surrounded by a spherical \textit{event horizon}, through which no form of matter or energy can pass outwards, not even light [10, 11]. For a Schwarzschild black hole of mass $M$ the radius of its horizon or \textit{Schwarzschild radius} is calculated as $R_{S}=2GM/c^{2}$, where, in the International System of Units (SI), $G=6.67 \times 10^{-11} N\cdot m^{2}\cdot kg^{-2}$ is the gravitational constant and $c=3\times 10^{8} m\cdot s^{-1}$ is the speed of light in vacuum [11]. From here on, we will mainly use SI units to express the values of the physical parameters involved.

\begin{figure}[H]
  \centering
    \includegraphics[width=0.35\textwidth]{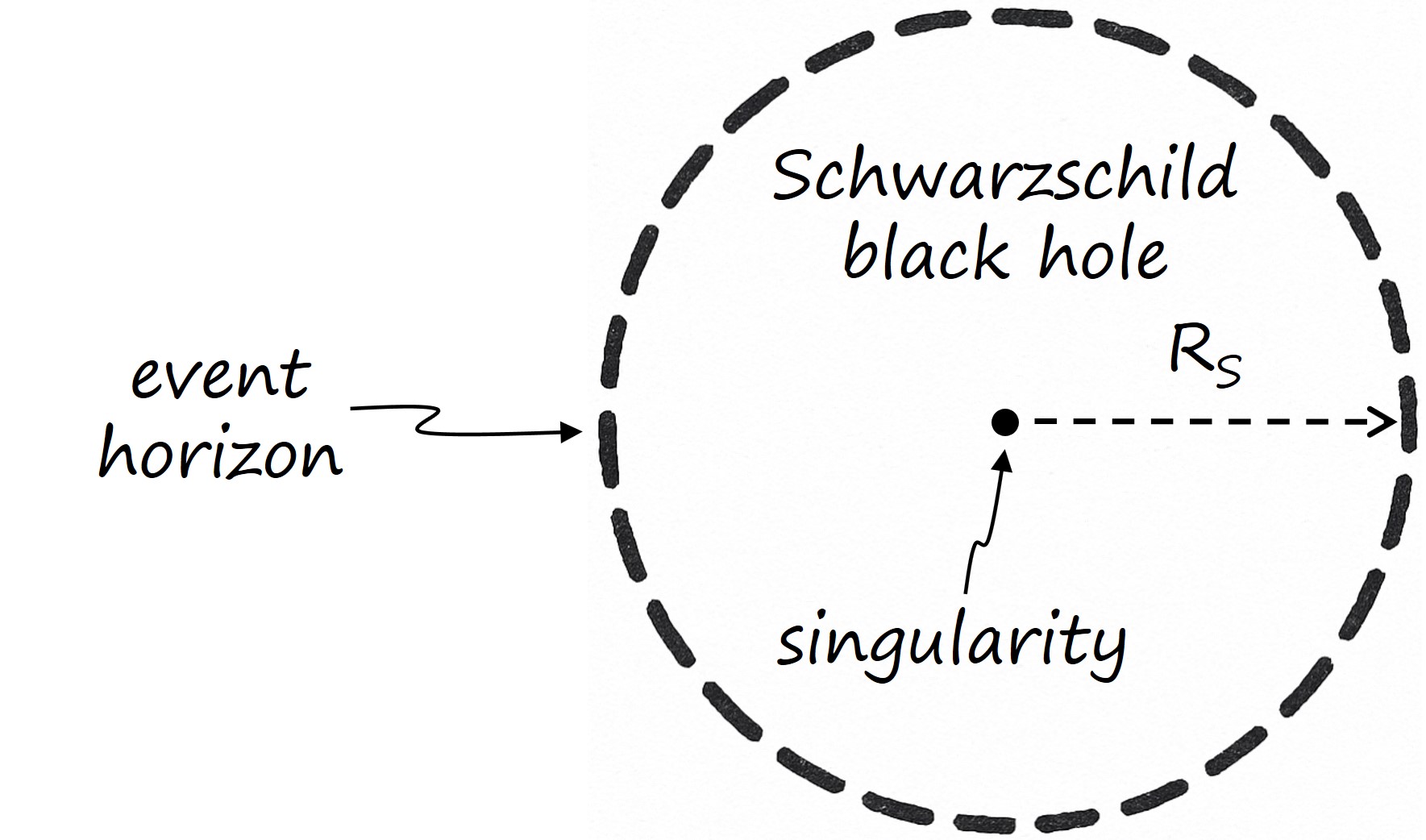}
  \caption{Internal structure of a Schwarzschild black hole.}
\end{figure}

In 1972, Hawking proved a result known as the \textit{area theorem}, which states that the area of the event horizon can only remain constant or increase [12]. The increase occurs, for example, when the black hole absorbs material from its surroundings. To understand this result, let us note that, according to the equation for the Schwarzschild radius, the area of the event horizon is given by

\begin{equation} 
A=4 \pi R_{S}^{2}=\frac{16\pi G^{2}M^{2}}{c^{4}}.
\end{equation}

We see then that when matter or energy crosses the horizon into the interior, there is an increase in $M$ and, consequently, in $A$. Hawking realized that his theorem had a remarkable similarity with the second law of thermodynamics, since the role of $A$ is analogous to that of entropy, which in an isolated physical system can only increase or remain constant. However, while Hawking thought that it was only a formal analogy between area and entropy, Bekenstein took the theorem seriously, proposing that a Schwarzschild black hole has an entropy directly proportional to $A$ [5,13]. According to Bekenstein, the entropy of the matter that enters the horizon, instead of disappearing, increases $A$, increasing the entropy of the black hole and decreasing the entropy of the universe in the exact proportion to maintain the second law [1].\\

Fig. 2 illustrates Bekenstein’s idea, where $A$ has been divided into a set of elementary cells of area $l_{P}^{2}$, where $l_{P}$ is the \textit{Planck length}, which is the smallest distance to which a physical meaning can be assigned [14]. The Planck length is defined as $l_{P}= (\hbar G/c^{3})^{1/2}=1.62\times 10^{-35} m$, where $\hbar =1.05 \times 10 ^{-34}J \cdot s$ is the reduced Planck constant. By defining each elementary area to have a size of $l_{P}^{2}$, Bekenstein ensured that it had the smallest possible value. This means that the total number of Planck areas contained within the horizon is, in general, a huge number, and is calculated as:

\begin{equation} 
N=\frac{A}{l_{P}^{2}}.
\end{equation}

\begin{figure}[H]
  \centering
    \includegraphics[width=0.35\textwidth]{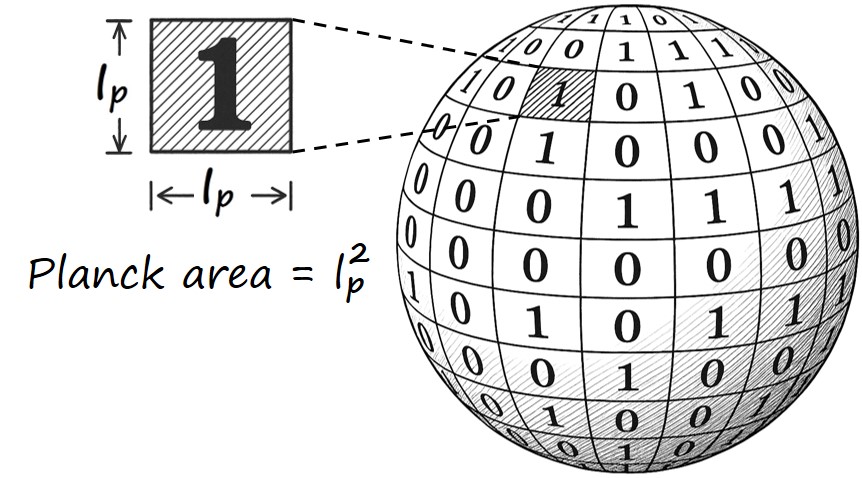}
  \caption{The event horizon can be imagined as a spherical surface of total area $A$ composed of a large number of elementary cells of area $l_{P}^{2}$, each one of which stores a bit of information (0 or 1).}
\end{figure}

To obtain the BH entropy from these ideas, let us recall that, in its microscopic formulation, the entropy $S$ of a physical system is determined by the number $\Omega$ of microscopic configurations or microstates that are compatible with a given macrostate [15,16]:

\begin{equation} 
S=k \ln \Omega,
\end{equation}

where $k=1.38 \times 10^{-23} J\cdot K^{-1}$ is the Boltzmann constant. According to Bekenstein’s proposal illustrated in Fig. 2, the macrostate of the Schwarzschild black hole is its mass $M$, while the microstates are bits of information stored in the surface cells of size $l_{P}^{2}$, each of which can represent one of two discrete values: 0 or 1. The total number of microstates encoded in the horizon area is $\Omega = 2^{N}$ which, by Eq. (2), can also be written as $\Omega = 2^{A/l_{P}^{2}}=2^{Ac^{3}/\hbar G}$, so that, by Eq. (3), the BH entropy is given by

\begin{equation}
S_{BH}=k \ln 2^{A/l_{P}^{2}}=\frac{kc^{3}A}{\hbar G} \ln 2 \cong 0.7 \frac{kc^{3}A}{\hbar G}.
\end{equation}

The value found by Bekenstein using much more complex reasoning is [1]: $S_{BH}=0.27 kc^{3}A/ \hbar G$. This expression differs from Eq. (4) only by a dimensionless constant. The exact expression found by Hawking for the entropy of a Schwarzschild black hole is very close to Bekenstein's result [6]: $S_{BH}=0.25kc^{3}A/ \hbar G= kc^{3}A/ 4\hbar G$. By introducing into this last equality the value of $A$ given by Eq. (2), we can also write the entropy as a function of the black hole mass $M$ and the solar mass $M_{\odot} = 1.99 \times 10 ^{30}kg$: 

\begin{equation} 
S_{BH}= \frac{4\pi kGM^{2}}{\hbar c} = 2.65 \times 10^{16}k \left( \frac{M}{kg} \right)^{2}=1.05 \times 10^{77}k \left( \frac{M}{M_{\odot}} \right)^{2},
\end{equation}

where we have introduced the values of the constants. By the general definition of entropy, Eq. (3), we obtain $\Omega = e ^{S_{BH}/k}$. The least massive black holes observed are stellar ones, whose masses are on the order of the solar mass, $\sim 10^{30}kg$. This number allows us to calculate a lower bound for $S_{BH}$ and $\Omega$. Thus, by Eq. (5) $S_{BH} \sim 10^{77}k$ so that $\Omega \sim e ^{10^{77}}$. No object with the same mass $M$ confined to a region of fixed size has an entropy greater than this. In other words, for a given total mass-energy, the physical state with the maximum possible entropy in the universe is a black hole.\\ 

We have reached the frontier of knowledge, as the physical meaning of $\Omega$ is unknown, nor is there any way to explain its colossal value, which contradicts the classical view of a black hole as a very simple object. The challenge of unraveling these enigmas remains in the hands of future generations of physicists.

\section{Hawking Temperature and Bekenstein-Hawking Entropy}

The most important consequence of the Hawking entropy is that black holes possess a temperature. Let us see how this result follows from the black-hole entropy in a simple way. \\

According to the first law of thermodynamics, the relation between the internal energy $E$, pressure $P$, volume $V$, absolute temperature $T$, and entropy $S$ of a system is $dE=TdS-PdV$ [15]. The event horizon, however, has no material existence; it merely represents a point of no return for matter and radiation crossing it inward. This means that a Schwarzschild black hole exerts no pressure on its surroundings, so $P=0$, and the first law reduces to $dE=TdS$. On the other hand, since the only parameter that defines a Schwarzschild black hole is its mass, by Einstein's mass-energy equivalence, a black hole of mass $M$ has a total internal energy $E=Mc^{2}$. Consequently, its absolute temperature is related to its entropy $S=S_{BH}$ as

\begin{equation}
\frac{1}{T}=\frac{d}{dE}S_{BH}=\frac{1}{c^{2}}\frac{d}{dM}S_{BH}.
\end{equation}

If we introduce Eq. (5) into this equality, we obtain

\begin{equation}
\frac{1}{T}=\frac{1}{c^{2}}\frac{d}{dM}\left( \frac{4\pi kGM^{2}}{\hbar c} \right)=\frac{8\pi kGM}{\hbar c^{3}}.
\end{equation}

Therefore, the temperature $T=T_{H}$ of the black hole is

\begin{equation} 
T_{H}=\frac{\hbar c^{3}}{8\pi kGM} \cong 1.23 \times 10^{23}K \left( \frac{kg}{M}\right)\cong 6.17 \times 10^{-8}K \left( \frac{M_{\odot}}{M}\right).
\end{equation}

This equation is known as the \textit{Hawking temperature}, and was derived by Hawking shortly after Bekenstein's discovery [13]. However, Hawking used a much more complex and detailed reasoning, which takes into account quantum effects near the event horizon [2,3,11].\\ 

\begin{figure}[H]
  \centering
    \includegraphics[width=0.6\textwidth]{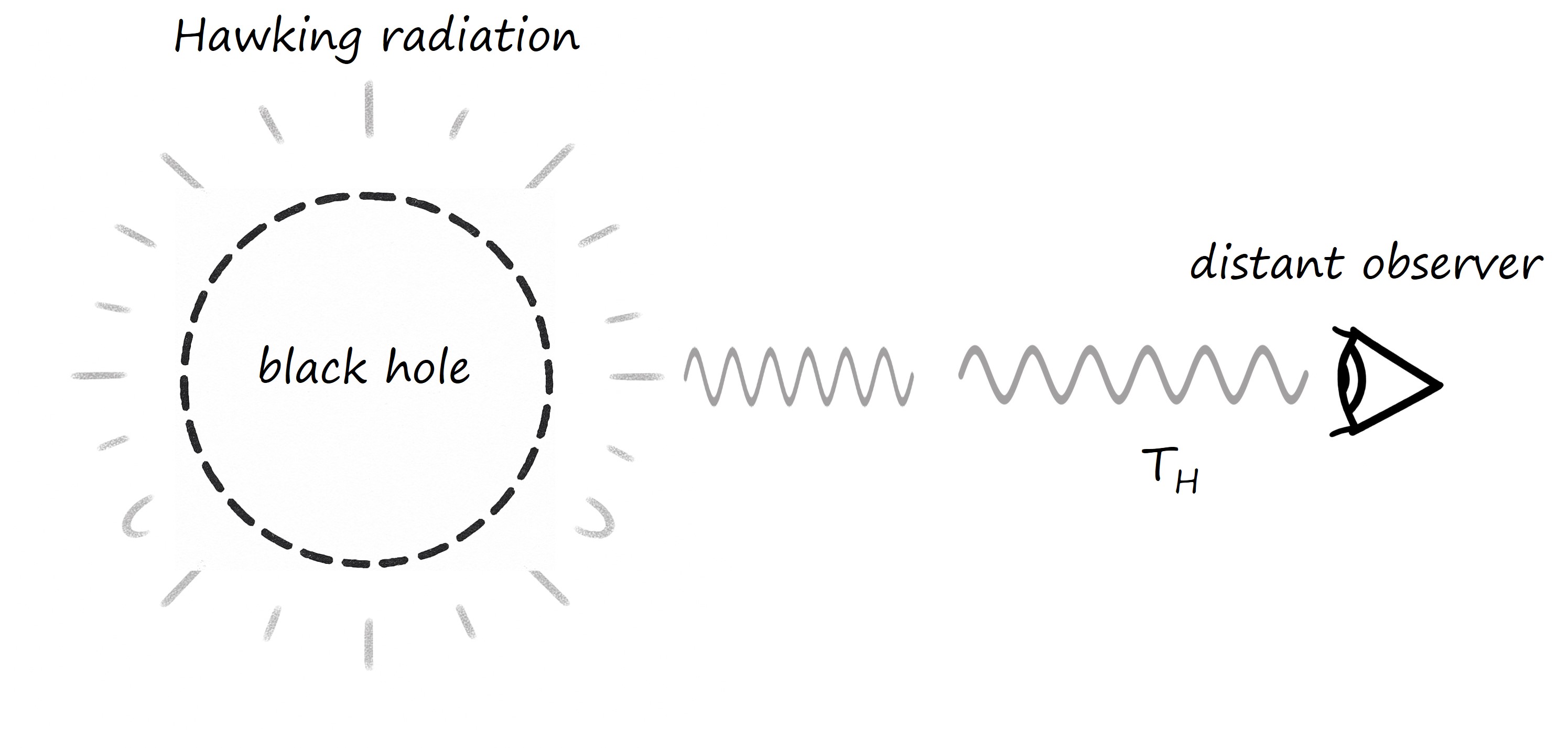}
  \caption{Hawking radiation causes the black hole to evaporate. A distant observer detects this radiation as thermal emission with a temperature $T_{H}$.}
\end{figure}

Since a body with a non-zero absolute temperature emits thermal radiation, Eq. (8) implies, on the one hand, that black holes are not black, and, on the other, that they evaporate, as the emitted radiation gradually carries away the mass-energy of the black hole to the outside (see Fig. 3). A distant observer detects Hawking radiation as thermal emission with an absolute temperature $T_{H}$ given by Eq. (8). \\ 

As we know, the least massive astrophysical black holes created by stellar collapse have a mass $M\sim M_{\odot} \sim 10^{30}kg$. This number allows us to calculate an upper bound for $T_{H}$, which, according to Eq. (8), corresponds to $T_{H} \sim 10^{-8}K$. This temperature is extremely small, and therefore undetectable by astronomical observations, which makes empirical verification of Bekenstein and Hawking's findings very difficult. This has led specialists to propose alternative ways to subject these findings to the verdict of experimentation [9,17].

\section{Concluding remarks}

We have seen that Bekenstein’s discovery not only challenges the classical view of a black hole as a very simple object, but also contradicts the very definition of a black hole. As is often the case with many revolutionaries, Bekenstein was not fully aware of the implications of his work, and it took the intervention of Hawking and others to turn it into a true scientific revolution. This work represents the first step toward the most ambitious goal in physics: developing a theory of quantum gravity that reconciles quantum mechanics and general relativity.\\ 

Bekenstein died in 2015, at the age of 68, and remained highly active, conducting research across various topics and disseminating his ideas. Given the importance and significance of his scientific contributions, Bekenstein will be remembered as the father of black hole thermodynamics, and his discoveries will undoubtedly continue to inspire the work of those seeking to unravel the great mysteries of the universe.

\section*{Acknowledgments}
I would like to thank to Daniela Balieiro for their valuable comments in the writing of this paper. 

\section*{References}

[1] J.D. Bekenstein, Black Holes and Entropy, Physical Review D 7 (1973) 2333–2346.

\vspace{2mm}

[2] S.W. Hawking, Black Hole explosions?, Nature 248 (1974) 30–31.

\vspace{2mm}

[3] S.W. Hawking, Particle creation by black holes, Communications in Mathematical Physics 43 (1975) 199–220.

\vspace{2mm}

[4] S.W. Hawking, The quantum mechanics of black holes, Scientific American 236 (1976) 34–40.

\vspace{2mm}

[5] S.W. Hawking, A brief history of time, Bantam Books, New York, 1998.

\vspace{2mm}

[6] S.W. Hawking, The Universe in a Nutshell, Bantam Books, New York, 2001.

\vspace{2mm}

[7] M.C. LoPresto, Some Simple Black Hole Thermodynamics, The Physics Teacher 41 (2003) 299–301.

\vspace{2mm}

[8] J. Pinochet, “Black holes ain’t so black”: An introduction to the great discoveries of Stephen Hawking, Phys. Educ. 54 (2019) 035014.

\vspace{2mm}

[9] J. Pinochet, Hawking for everyone: commemorating half a century of an unfinished scientific revolution, Phys. Educ. 59 (2024) 055001. https://doi.org/10.1088/1361-6552/ad589c.

\vspace{2mm}

[10] J.P. Luminet, Black Holes, Cambridge University Press, Cambridge, 1999.

\vspace{2mm}

[11] B. Schutz, Gravity from the Ground Up, Cambridge University Press, Cambridge, 2003.

\vspace{2mm}

[12] S.W. Hawking, Black Holes in General Relativity, Communications in Mathematical Physics 25 (1972) 152–166.

\vspace{2mm}

[13] K. Ferguson, Stephen Hawking. Su vida y obra, Crítica, Buenos Aires, 2012.

\vspace{2mm}

[14] R.J. Adler, Six easy roads to the Planck scale, American Journal of Physics 78 (2010) 925–932.

\vspace{2mm}

[15] F. Reif, Fundamentals of Statistical and Thermal Physics, Waveland Press, Illinois, 2009.

\vspace{2mm}

[16] R.A. Serway, J.W. Jewett, Physics for Scientists and Engineers with Modern Physics, Thomson, Cengage Learning, 2010.

\vspace{2mm}

[17] J. Steinhauer, Observation of self-amplifying Hawking radiation in an analogue black-hole laser, Nature Physics 10 (2014) 864–869.

\end{document}